\documentclass[aps,pre,twocolumn,showpacs,floatfix]{revtex4}
\usepackage{graphicx}
\usepackage{array}
\usepackage[usenames]{color}
\usepackage{amsmath}
\usepackage{ulem}
\newcommand{\tens}[1]{\,\raisebox{0ex}[0ex][0ex]{\uuline{\mbox{$#1$}}}\,}
\newcommand{\Tr}{\ensuremath{\operatorname{Tr}}}
\begin{document}  
\newcommand{\eigena}{a}
\newcommand{\eigenb}{b}
\newcommand{\auto}[4]{\ensuremath{\left<#1\left(#2\right)#3\left(#4\right)\right>}}

\title{Conjugate gradient heatbath for ill-conditioned actions}
\author{Michele Ceriotti}\email{michele.ceriotti@phys.chem.ethz.ch}
\author{Giovanni Bussi and Michele Parrinello}
\affiliation{Computational Science, Department of Chemistry and Applied Biosciences,
ETH Zurich, USI Campus, Via Giuseppe Buffi 13, CH-6900 Lugano, Switzerland}
\begin{abstract}
We present a method for performing sampling from a Boltzmann distribution
of an ill-conditioned quadratic action.
This method is based on heatbath 
thermalization along a set of conjugate directions, generated via 
a conjugate-gradient procedure. The resulting scheme outperforms 
local updates for matrices with very high condition number, since it avoids
the slowing down of modes with lower eigenvalue, and 
has some advantages over the global heatbath approach, compared to 
which it is more stable and allows for more freedom in devising 
case-specific optimizations.
\end{abstract}
\maketitle

A common problem in many branches of statistical physics is the sampling of 
distributions of the type $p\propto \exp\left(-\frac{1}{2}{\bf x\tens{A} x}\right)$
where $\tens{\bf A}$ is a positive definite $N\times N$ matrix and the
random variable ${\bf x}$ an $N$-dimensional vector.
Areas in which such sampling is needed are for instance QCD\cite{lusc94npb,divi+95npb,forc99parc} 
and a recently developed linear scaling electronic structure 
method\cite{kraj-parr05prb,kraj-parr06prb}.
In principle sampling $p$ is straightforward, if diagonalizing $\tens{\bf A}$ is
an option. However, in many cases, $N$ is so large that circumventing the 
$\mathcal{O}\left(N^3\right)$ diagonalization step becomes mandatory.
Different approaches have been proposed. In the so-called global heatbath
method  one writes $\tens{\bf A}=\tens{\bf M}^T\tens{\bf M}$,
and obtains a series of statistically independent vectors by solving  the linear
system $\tens{\bf M}{\bf x}={\bf R}$, where ${\bf R}$ is a vector whose components
are distributed according to a Gaussian with zero mean and 
unit variance $\left<R^2\right>=1$. The advantage of this method is that
the algorithmic complexity of the problem can be reduced
by using an iterative solver for the linear system.
 In order to expedite sampling a Metropolis-like
criterion has been suggested that leads to correct sampling
without having to bring the iterative process to full convergence\cite{forc99pre,wilc02npb}.
Unfortunately, when the ratio between the largest and smallest eigenvalues is large 
(ill-conditioned matrices) the acceptance of 
this scheme drops to zero unless full convergency is achieved.
An alternative approach is the local heatbath algorithm, in which 
at every step one single component of the state vector ${\bf x}$ 
is thermalized in turn, keeping the others fixed.
It has been pointed out elsewhere\cite{good-soka89prd,adle89npb} that there is a 
close analogy between this second method and the Gauss-Seidel minimization technique.
This approach is relatively inexpensive, but becomes very inefficient when the 
condition number of $\tens{\bf A}$ is large, and even more inefficient when 
the observable of interest depends strongly on the eigenvectors  corresponding to smaller eigenvalues.

In this paper we propose a heatbath algorithm in which moves are performed along 
mutually conjugated
directions. This choice is based on the analogy between various heatbath methods 
(see e.g. Ref.~\cite{good-soka89prd}) and directional minimization techniques.
We show both analytically and numerically that the choice of conjugate directions
allows all the degrees of freedom to become  decorrelated on the same time scale, 
independent of their associated eigenvalue. We also discuss the cases in which
the improved efficiency outbalances the additional computational cost. 
Our method can be interpreted as the subdivision of the global heatbath matrix 
inversion process into $N$ intermediate steps, all of which guarantee an 
exact sampling of the probability distribution.

In section~\ref{sec:cartaepenna} we introduce a simple 
formalism to treat heatbath moves along general directions, 
discuss the properties of a sweep through
a set of conjugate directions, and describe a couple of 
algorithms to obtain such a set with reasonable effort. In 
section~\ref{sec:numerico} we present some numerical tests on 
a model action and compare the efficiency of conjugate
directions heatbath with local moves for a model observable.
In section~\ref{sec:comp-global} we compare our method with global
heatbath, and in section~\ref{sec:conclusions} we present our conclusions.

\section{\label{sec:cartaepenna}Collective modes heatbath}
Given a probability distribution
\begin{equation}
P({\bf x})\propto \exp\left[-\left(\frac{1}{2}{\bf x}\tens{\bf A}{\bf x}-{\bf b}\cdot{\bf x}\right)\right]
\end{equation}
a generic heatbath algorithm can be described as a stochastic process
in which the vector ${\bf x}\left(t+1\right)$ is related to the vector at 
the previous step ${\bf x}\left(t\right)$ by
\begin{equation}
{\bf x}\left(t+1\right)={\bf x}\left(t\right)+\tau{\bf d} \label{eq:step},
\end{equation}
where ${\bf d}$ is a direction in the ${\bf x}$ space and 
\begin{equation}
\tau = -\frac{{\bf d}\left(\tens{\bf A}{\bf x}-{\bf b}\right)}{{\bf d}\tens{\bf A}{\bf d}}+
\left(\beta {\bf d}\tens{\bf A}{\bf d}\right)^{-1/2} R \label{eq:hb-tau}
\end{equation}
where $R$ is a Gaussian random number with zero mean and unitary spread $\left<R^2\right>=1$, and $\beta$ is the inverse temperature at which the 
sampling is performed.
The application of this algorithm does not require inversion of the matrix
$\tens{\bf A}$.
The sequence of directions ${\bf d}$ is rather arbitrary, and could be
a random sequence or a predefined deterministic sequence.
Strictly speaking, detailed balance is satisfied only if
the directions are randomly chosen at each step.
Nevertheless it has been shown in Ref\cite{mano-deem99jcp} that correct sampling
can be achieved if every Monte Carlo move leaves the
equilibrium distribution unchanged. In Appendix~\ref{sec:stationary} we show
that this is the case, provided that direction ${\bf d}$ is chosen independently
from position ${\bf x}$. 
Nevertheless, different choices of directions can lead to different sampling efficiency.
Our final choice will be to select for ${\bf d}$ a sequence of conjugate directions 
(Section~\ref{sub:conj-dir}). However, we shall first analyze the choice
of random, uncorrelated directions, and a sequential sweep along a set of orthogonal
directions.

For the sake of simplicity, we take ${\bf b}=0$ and we choose the basis into 
which $\tens{\bf A}$ is diagonal, $A_{ij}=\eigena_i\delta_{ij}$.
Since these properties are subsequently never used, no loss of generality is implied.
To compare the efficiency of the different choices of directions we shall consider
the autocorrelation matrix for the components along the eigenmodes 
$\auto{x_i}{0}{x_j}{t}$. A quantitative measure of the speed of decorrelation of
$\auto{x_i}{0}{x_j}{t}$ can be obtained from its slope at the origin. Since in Monte Carlo
one progresses in discrete steps, this quantity is given by
\begin{equation}
\left<x_i\left(0\right) x_j\left(1\right)\right>=
\sqrt{\left<x_i\left(0\right)^2\right>\left<x_j\left(0\right)^2\right>}
\left[\delta_{ij}-\Delta_{ij}\left({\bf d}\right)\right]\label{eq:slopefirst}
\end{equation}
In Eq. (\ref{eq:slopefirst}) we have introduced the normalized slope 
tensor $\tens{\boldsymbol\Delta}$, which can 
be expressed as a function of the eigenvalues of $\tens{\bf A}$ and of the components
of ${\bf d}$, using equations (\ref{eq:step}) and (\ref{eq:hb-tau}):
\begin{equation}
\Delta_{ij}\left({\bf d}\right)=\frac{\eigena_i h_i h_j}{\sum_k \eigena_k h_k^2}
\frac{\left<x_i\left(0\right)^2\right>}
{\sqrt{\left<x_i\left(0\right)^2\right>\left<x_j\left(0\right)^2\right>}}=
\frac{\sqrt{\eigena_i\eigena_j} d_i d_j}{\sum_k \eigena_k d_k^2}
\label{eq:slope-tens}
\end{equation}
Therefore, depending on the choice of direction ${\bf d}$, the different
components of the vector ${\bf x}$ decorrelate at different speeds.
However, since $\Tr \tens{\boldsymbol\Delta}=1$, the sum of these
normalized speeds does not depend on the direction chosen.
The same quantity $\tens{\boldsymbol\Delta}$ also enters a recursion relation 
for the autocorrelation functions at a generic Monte Carlo step $t$, 
\begin{eqnarray}
\auto{x_i}{0}{x_j}{t+1}=\nonumber\\
\auto{x_i}{0}{x_j}{t}-
\sum_k\left[\auto{x_i}{0}{x_k}{t}\sqrt{\frac{\eigena_k}{\eigena_j}}
\Delta_{kj}\left({\bf d}\right)\right]\label{eq:auto-induction}
\end{eqnarray}
Use of this equation requires that one appropriately averages over the direction
${\bf d}$, as we shall discuss in the following.

We will begin our analysis from the simpler case, in which the direction ${\bf d}$
is chosen at every step to be equal to a stochastic vector ${\bf R}$, whose 
components are distributed as Gaussian random numbers with zero mean and 
standard deviation one. The normalized slope tensor (\ref{eq:slope-tens}) in this 
case results from an average over the possible directions, 
\begin{equation}
\left<\Delta_{ij}\left({\bf d}={\bf R}\right)\right>=
\eigena_i \delta_{ij} \left<\frac{R_i^2}{\sum_k\eigena_k R_k^2}\right>
\overset{N\rightarrow\infty}{\approx}\frac{\eigena_i \delta_{ij}}{\Tr\tens{\bf A}}.
\label{eq:random-slope}
\end{equation}
The limit expression holds for the size $N$ of the matrix going to infinity 
(see Appendix~\ref{sec:rnd-asymptotic}), 
under the hypothesis that the largest eigenvalue of $\tens{\bf A}$ does not grow
with $N$ and that $\Tr \tens{\bf A}$ is $\mathcal{O}\left(N\right)$, 
hypotheses which are relevant to many physical problems.
Since in this case the direction chosen at every step is independent of all
the previous choices, the same average enters equation (\ref{eq:auto-induction}) 
at any time, so that proceeding by induction one can easily obtain the 
entire autocorrelation function,
\begin{equation}
\auto{x_i}{0}{x_j}{t}=
\delta_{ij}\left<x_i\left(0\right)^2\right>
\left[1-\left<\Delta_{ij}\right>\right]^t \label{eq:random-full}
\end{equation}
where $\left<\Delta_{ij}\right>$ is the quantity obtained in 
equation~(\ref{eq:random-slope}). From (\ref{eq:random-full}) we can 
calculate the autocorrelation time for mode $i$, 
\[
\tau_i=\frac{\sum_{t=0}^\infty \auto{x_i}{0}{x_i}{t}}{\left<x_i^2\right>}
=\left[\eigena_i\left<\frac{R_i^2}{\sum_k\eigena_k R_k^2}\right>\right]^{-1}
\overset{N\rightarrow\infty}{\approx}
\frac{\Tr\tens{\bf A}}{\eigena_i}.
\]
In the case of large $N$, the decorrelation speed of the components along normal modes
is directly proportional to the corresponding eigenvalue, so that in ill-conditioned
cases a critical slowing down for the softer normal modes will be present.

Let us now consider moves along a predefined set of orthogonal directions 
$\left\{{\bf u}^{(m)}\right\}_{m=0\ldots N-1}$. This is done  to mimic 
the case in which one performs a sweep along Cartesian directions.
In our reference frame, where $\tens{\bf A}$ is taken to be diagonal, this
would be trivial, hence the choice of an arbitrarily oriented set of orthogonal 
directions. As in standard local heatbath, the outcome will depend on the
orientation of the $\left\{{\bf u}^{(m)}\right\}$ relative to the
eigenvectors of $\tens{\bf A}$.
Averaging over all the possible choices of initial direction, we find the
slope at $t=0$,
\begin{equation}
\left<\Delta_{ij}\right>=\frac{1}{N}\sum_m\Delta_{ij}\left({\bf u}^{(m)}\right)=
\frac{1}{N}\sum_m\frac{\sqrt{\eigena_i \eigena_j} u_i^{(m)}u_j^{(m)}}{\sum_k{\eigena_k u_k^{(m)}}^2}\label{eq:localdelta}
\end{equation}
Obviously, it is not possible to reduce this result to an expression which
does not depend on the particular set of orthogonal directions.
However, the following inequality holds
\begin{equation}
\frac{\eigena_i \delta_{ij}}{N\eigena_{max}} 
\le \left<\Delta_{ij}\right> \le
\frac{\eigena_i \delta_{ij}}{N\eigena_{min}}  \label{eq:localuneq}
\end{equation}
Equation~(\ref{eq:localuneq}) does not put rigid constraints on the 
value of $\left<\Delta_{ij}\right>$, but demonstrates that also in this case 
$\tens{\boldsymbol\Delta}$ is diagonal and
suggests that in real life the convergence will be faster for the higher eigenvalues,
and that the spread in the relaxation speed for different modes is 
larger when the condition number $\kappa = \eigena_{max}/\eigena_{min}$ is higher.

In the case where directions $\left\{{\bf u}^{(m)}\right\}$ are swept sequentially
we have not been able to derive a closed expression for $\auto{x_i}{0}{x_j}{t}$
because of the dependence of ${\bf d}\left(t\right)$ on the previous history.
If, on the other hand, a random direction is drawn from $\left\{{\bf u}^{(m)}\right\}$
at every step,  $\auto{x_i}{0}{x_j}{t}$ is given by expression (\ref{eq:random-full})
where $\left<\Delta_{ij}\right>$ has the value in equation~(\ref{eq:localdelta}).

\subsection{\label{sub:conj-dir}Moves along conjugate directions}

It is clear from equation (\ref{eq:random-full}) that a random choice of the directions
${\bf d}$ leads to fast decorrelation of the components relative to the eigenvectors 
with high eigenvalues. On the other hand, the components relative to the eigenvectors with 
low eigenvalues will decorrelate more slowly. 
Similar behavior is expected for the local heatbath method, unless particular
relations hold between the eigenvectors and the Cartesian axes. 
If the operator $\tens{\bf A}$ is ill-conditioned, the practical
consequence is that the slow modes will be accurately sampled only after a very large
number of steps.
As we have already discussed, the sum of the decorrelation slopes of the different
components does not depend on the choice of the directions ${\bf d}$.
However, with a proper choice of the directions ${\bf d}$ this sum could
be spread in a uniform way among the different modes.
A similar problem arises in minimization algorithms based on directional search,
and is often solved choosing a sequence of conjugated directions\cite{numerical-recipes}.
In the same spirit, we can compute the decorrelation speed of the different
modes when the ${\bf d}$'s are chosen to be conjugated directions.
Let us consider a set of conjugated directions $\left\{{\bf h}^{(i)}\right\}$,
such that ${\bf h}^{(i)}\tens{\bf A}{\bf h}^{(j)}=\delta_{ij}$. 
The set $\left\{{\bf h}^{(i)}\right\}$
can be generated with various algorithms, such as a Gram-Schmidt 
orthogonalization that uses the positive definite $\tens{\bf A}$ matrix as a metric,
or a conjugate gradient procedure, as described in Section~\ref{sub:tricks}.

Using the fact that
$\sum_k h_i^{(k)}h_j^{(k)}=\eigena_i^{-1}\delta_{ij}$,
the slope at $t=0$ is
\begin{align*}
\left<\Delta_{ij}\right>=\frac{1}{N}\sum_m
\frac{\sqrt{\eigena_i \eigena_j} h^{(m)}_i h^{(m)}_j}{{\bf h}^{(m)}\tens{\bf A}{\bf h}^{(m)}}=\\
=\frac{1}{N}\sqrt{\frac{\eigena_j}{\eigena_i}}\sum_m
\frac{\eigena_i h^{(m)}_i h^{(m)}_j}{{\bf h}^{(m)}\tens{\bf A}{\bf h}^{(m)}}=\frac{\delta_{ij}}{N}
\end{align*}
With this choice, the decorrelation slopes of the different
modes are independent of the eigenvalue.
If one chooses one conjugate direction at random at each step it is 
straightforward to show that overall the autocorrelation function decays 
exponentially as
\[
\left<x_i\left(0\right) x_j\left(t\right)\right>=
\delta_{ij}\left<x_i\left(0\right)^2\right>
\left[1-\frac{1}{N}\right]^t
\]

This derivation shows that if matrix $\tens{\bf A}$ is ill-conditioned and
one wishes to decorrelate the slow modes, then the choice of performing
the heatbath using a sequence of conjugated directions can improve
the sampling quality dramatically. Of course, the slow modes are accelerated
and the fast modes are decelerated.
However, it is clear that a completely independent vector ${\bf x}$ is
obtained only when all the modes are decorrelated.
A heatbath on conjugate directions allows all the modes to be decorrelated
with the same efficiency, irrespective of their stiffness.
Even better efficiency can be obtained by sequentially sweeping a set of
conjugated directions. At first sight it would appear that the dependence
of ${\bf h}\left(t\right)$ on ${\bf h}\left(t-1\right)$ would make it
very difficult if not impossible to obtain the autocorrelation function 
in a closed form.
However, conjugate directions have a redeeming feature.
If we expand the position vector on the non-orthogonal basis 
$\left\{{\bf h}^{(m)}\right\}$, ${\bf x}=\sum_i \alpha^i {\bf h}^{(i)}$,
and we evaluate the correlation matrix between
the contravariant components $\alpha^i$, we find that 
$\left<\alpha^i\alpha^j\right>=\delta_{ij}$. 
This property can be easily demonstrated taking into account that the
ensemble average $\left<x_i x_j\right>=A^{-1}_{ij}$,
and that conjugacy implies
${\bf h}^{(i)}\tens{\bf A}{\bf h}^{(j)}=\delta_{ij}$.
Thus, effectively, every time we perform a heatbath move
along direction ${\bf h}^{(i)}$ the component $\alpha^i$ is randomized,
without affecting the others. After a complete sweep across the set of directions
a completely independent state is obtained. 

A more formal proof is provided in appendix~\ref{sec:cd-formal}, where it is
also demonstrated that the autocorrelation function is
\begin{equation}
\left<x_i\left(0\right) x_i\left(t\right)\right>=
\left<x_i\left(0\right)^2\right>
\left\{
\begin{array}{cc}
\left[1-\frac{t}{N}\right] 	& t<N\\
0				& t\ge N
\end{array}
\right.\label{eq:cd-autofun}
\end{equation}
Therefore the corresponding autocorrelation time is $\tau_i=\left(N+1\right)/2$.
A remarkable feature of equation (\ref{eq:cd-autofun}) is that the
autocorrelation function is linear, and that after $N$ moves 
a completely independent vector is obtained. 
This property holds also for the global heatbath method. 
In Section~\ref{sec:comp-global} we shall discuss the relation between
our approach and  global heatbath sampling.

\subsection{\label{sub:tricks}Conjugate-gradient approach to generate conjugate directions}
In the last section we have shown how a heatbath algorithm based on
conjugate directions can dramatically improve the sampling of the slow modes
for an ill-conditioned action. 
An efficient strategy to generate these directions is the application of the 
conjugate gradient procedure\cite{numerical-recipes}.  
For the sake of completeness and to introduce a consistent notation we give 
here an outline of the CG algorithm.
One starts from a random configuration and search direction, 
${\bf h}^{(0)}={\bf g}^{(0)}={\bf R}$, so that the directions obtained and 
the sample vector ${\bf x}$ are independent as required. Then, 
a series of directions ${\bf h}^{(m)}$ and residuals ${\bf g}^{(m)}$ are 
generated using the recurrence relations 
\[ 
{\bf g}^{(i+1)}={\bf g}^{(i)}-\lambda_i \tens{\bf A}\cdot{\bf h}^{(i)}\quad
{\bf h}^{(i+1)}={\bf g}^{(i+1)}+\gamma_i \cdot{\bf h}^{(i)}
\]
\[
\lambda_i=\frac{{\bf g}^{(i)}\cdot {\bf g}^{(i)}}{{\bf h}^{(i)}\tens{\bf A}{\bf h}^{(i)}}
\quad 
\gamma_i=\frac{{\bf g}^{(i+1)}\cdot {\bf g}^{(i+1)}}{{\bf g}^{(i)}\cdot {\bf g}^{(i)}}
\]
This procedure generates at every step a new direction ${\bf h}^{(i)}$, conjugated to all
the previous ones, and it can be used to perform a directional heatbath move on
${\bf x}$. It should be stressed that
there is no need to store all the  ${\bf h}^{(i)}$ if the heatbath moves 
are performed concurrently with the CG minimization.
The ``force'' $\tens{\bf A}{\bf h}^{(i)}$ can be reused for performing 
 the heatbath update (cfr. Eq. (\ref{eq:step})).
At a certain point the CG procedure will be over, with the residual ${\bf g}$ 
dropping to zero. The sequential sweep algorithm described inte the 
previous section can be implemented starting again from the same 
${\bf g}^{(0)}$.

In contrast to the global heatbath method, numerical stability is not 
a major issue, since the accuracy of the sampling does not depend
on the search directions being exactly conjugated. The only effect 
of imperfect conjugation would be to slightly reduce 
 the decorrelation efficiency.
There is however a  drawback to this approach. In order to be ergodic, 
the set of directions must span the whole space. 
The problem arises when there are degenerate 
eigenvalues, as CG converges to zero in a number $p$ of iterations equal to the
number of distinct eigenvalues. If we keep reusing the same set of $p<N$ 
directions, only a part of the subspaces corresponding to degenerate eigenvalues
will be explored, and the sampling will not be ergodic.

\begin{figure}
\caption{\label{fig:hybrid}Scheme of the block algorithm 
described in paragraph \ref{sub:tricks}; squares represent 
eigenvectors of the action matrix, which need to be refreshed in order
to obtain a statistically independent sample point; modes on 
the same column correspond to the same, degenerate eigenvalue.
At every step, one of the vectors of a set with the same 
size as the biggest degenerate subspace is used in a conjugate
gradient minimization, while the remaining ones are made 
orthogonal to the search directions that are generated in 
the process. When the first vector approaches zero, one can start
back on the second one (Figure~b)), and the process can be 
continued (Figures~c) and d)) until the refresh is complete.
}
\includegraphics[width=0.9\columnwidth]{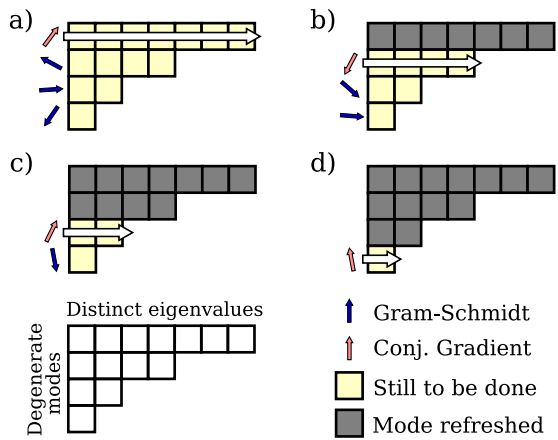}
\end{figure}

We have considered two possible ways of recovering ergodicity.
The simplest consists in 
drawing a new random point ${\bf g}^{(0)}={\bf R}$ every time 
we reset the CG search. 
This causes a deviation from the linear behavior of the autocorrelation
functions for $t\approx N$.
Non-degenerate eigenvalues will initially converge with $-1/p$ instead of 
$-1/N$ slope, but degenerate ones will converge more slowly, and with exponential
trend, as we are sampling random directions within every degenerate 
subspace.

In order to improve the efficiency, we mix CG with Gram-Schmidt
orthogonalization of a small set of vectors, ideally of the same size $d$ 
of the largest degeneracy present. 
As discussed earlier, here Gram-Schmidt 
orthogonalization has to be performed using the metric of $\tens{\bf A}$,
which amounts to imposing conjugacy.
The procedure is illustrated in 
Figure~\ref{fig:hybrid}. 
We start from $d$ random vectors, $\left\{{\bf v}^{(j)}\right\}_{j=0..d-1}$. We
set  ${\bf h}^{(0)}={\bf g}^{(0)}={\bf v}^{(0)}$ and begin a CG minimization.
At each step we obtain a search direction ${\bf h}^{(i)}$, and make each 
of the other $d-1$ vectors conjugate to ${\bf h}^{(i)}$ with 
a Gram-Schmidt procedure. 
This does not require any matrix-vector product other than the one necessary 
for the heatbath step. 
After $p$ iterations the conjugate gradient will have converged and ${\bf g}$ 
will be close to zero. We can start again
from the second vector in the pool, which meanwhile has 
become $\bar{\bf v}^{(1)}$, and is conjugate to all the directions visited so far. 
Thus, we  set ${\bf h}^{(0)}={\bf g}^{(0)}=\bar{\bf v}^{(1)}$ 
and start again the CG procedure, orthogonalizing the $d-2$ remaining vectors 
to ${\bf h}^{(i)}$, and so on and so forth.
After $N$ steps the procedure will be converged. At the successive sweep, one
can generate again a set of random initial $\left\{{\bf v}^{(j)}\right\}$. This
can make the method more stable, at the cost of some loss in performance.
Some savings can be made if one stores the conjugated $\bar{\bf v}^{(i)}$, 
and uses them in the subsequent sweeps, avoiding the need to repeat the GS 
orthogonalizations (see figure~\ref{fig:hybrid}).
In practice, where more than one complete sweep is affordable, it is easy to devise
adaptive variations of this scheme, in which the pool of vectors  
$\left\{{\bf v}^{(j)}\right\}$ is enlarged whenever the CG minimization converges 
in less than $N$ steps, so that in a few sweeps the optimal size to guarantee
ergodicity is attained.

\section{\label{sec:numerico}Benchmarks and comparison with local heatbath}
\begin{figure}
\caption{\label{fig:curve}Autocorrelation functions for 
a) the projection along the mode $\eigena_0=1$;
b) the projection along the mode $\eigena_4\approx 9.8$
for a matrix of the form (\ref{eq:mat-phonon})
with $N=100$ and condition number $\kappa=10^3$.
Line {\bf A} corresponds to local heatbath moves (one step 
stands for a complete sweep of the $N$ coordinates), lines
{\bf B} to {\bf D} to conjugate directions moves:
{\bf B} is the hybrid conjugate gradient/Gram-Schmidt block algorithm;
{\bf C} corresponds to CG sweeps, with the search direction 
randomized at the beginning of every sweep; 
curve {\bf D} corresponds to CG sweeps starting from the same 
initial vector. 
Conjugate direction moves decorrelate faster than local heatbath for the slow mode, but are
less efficient for modes with higher eigenvalue. For degenerate 
eigenmodes, the method used for curve {\bf D} is not ergodic (and thus
gives incorrect values for $\left<x_i^2\right>$), and
random restarts (curve {\bf C}) are much less efficient than the hybrid (curve {\bf B})
algorithm.
}
{\centering
\includegraphics[width=0.9\columnwidth]{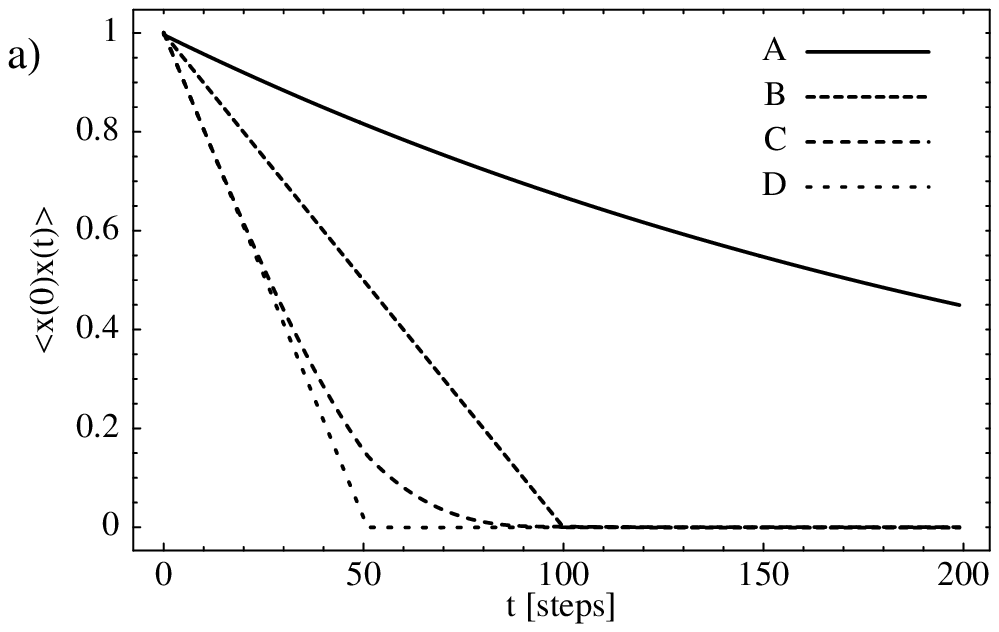}\\~\\
\includegraphics[width=0.9\columnwidth]{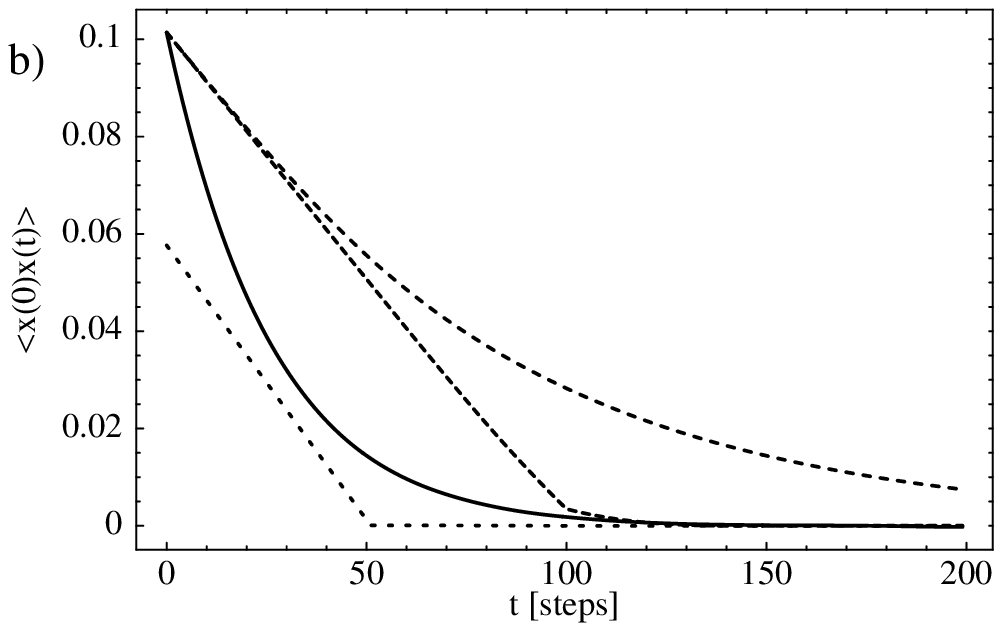}
}
\end{figure}

In the previous section we have discussed a collective modes heatbath method
that could outperform standard local heatbath techniques when
the Hamiltonian has a very large condition number and 
sampling along the slower eigenmodes is required.
In this section we illustrate the efficiency of our
algorithm using numerical experiments on a simple
model for $\tens{\bf A}$,
\begin{equation}
\tens{\bf A}=\tens{\bf 1}+\left(\begin{array}{cccccc}
-2b 	& b	& 0	&\cdots	& 0	& b	\\
b	&-2b 	& b	& 0	&\cdots	& 0	\\
0	& b 	&-2b	& b	&\ddots 	&\vdots	\\
\vdots	& 0	& b	&-2b 	&\ddots	& 0	\\
0	& \vdots& \ddots&\ddots	&\ddots	& b	\\
b	& 0	&\cdots	& 0	& b	&-2b	
\end{array}\right)
\label{eq:mat-phonon}
\end{equation}
This matrix corresponds to the dynamical matrix of a linear chain of 
spring-connected masses, with periodic
boundary conditions and an additional diagonal term to make  the acoustic mode
nonzero. $b$ can be chosen so as to obtain the desired condition number.
Eigenmodes and eigenvalues for such a matrix are easily obtained, 
\[
\eigena_k=1+2b\left(1-\cos\frac{2k\pi}{N}\right)
\]
\[
u^{(k)}_l=\sqrt{\frac{1+\delta_{0k}+\delta_{N/2,k}}{N}}
\left\{
\begin{array}{lc}
\cos\frac{2kl\pi}{N} & k\le N/2 \\
\sin\frac{2kl\pi}{N} & k > N/2
\end{array}
\right.
\]
and projection of a state on the eigenvectors is quickly 
done via fast-Fourier transform.
In Figure~\ref{fig:curve} we compare the the autocorrelation functions 
obtained with different algorithms for a matrix of the form (\ref{eq:mat-phonon}).
Figure~\ref{fig:curve} also highlights the ergodicity problems connected with the 
naive use of the conjugate gradient algorithm to generate the search directions, 
and shows how both the suggestions of paragraph~\ref{sub:tricks} can help in
solving this problem. 
In general, a conjugate directions search speeds up
decorrelation for the slower modes, but is less efficient than local heatbath 
for the modes with a high eigenvalue. This is a direct consequence of
the fact that $\Tr\tens{\boldsymbol{\Delta}}=1$. An additional advantage of 
our method is the linear rate of decorrelation, which allows complete 
decorrelation just like the direct inversion of $\tens{\bf M}$, whereas moves along the 
Cartesian axes lead to approximatively exponential autocorrelation functions.

We stress again that the relative efficiency of the two methods
depends strongly on the observable being calculated and on the
actual spectrum of the Hamiltonian of the system.
As a more realistic benchmark we will consider the evaluation of
the trace of the inverse matrix, i.e.
\begin{equation}
\Omega=\Tr \left(\tens{\bf A}^{-1}\right)=\left<{\bf x}^2\right> 
\label{eq:omega}
\end{equation}
This observable is strongly dependent on the slow modes.

\begin{figure}
\caption{\label{fig:benchmark}(Color online) Comparison of the 
efficiency of local heatbath versus conjugate-gradient moves.
The graph represents $\tau_{CG}/\tau_{loc}$,
the ratio of the autocorrelation times for the observable $\Omega$ (\ref{eq:omega}); 
$\tau_{loc}$ corresponds to the value obtained from standard local heatbath moves 
(one unit of
Monte Carlo time corresponds to a whole coordinates sweep), while
$\tau_{CG}$ corresponds to the value obtained with moves along 
conjugate directions, as obtained from our block algorithm with random restarts.
The data plotted results from a linear interpolation of some
simulations (labeled by $\otimes$) performed for an 
action of the form (\ref{eq:mat-phonon}), 
with varying size $N$ and condition number $\kappa$.
}
\includegraphics[width=0.9\columnwidth]{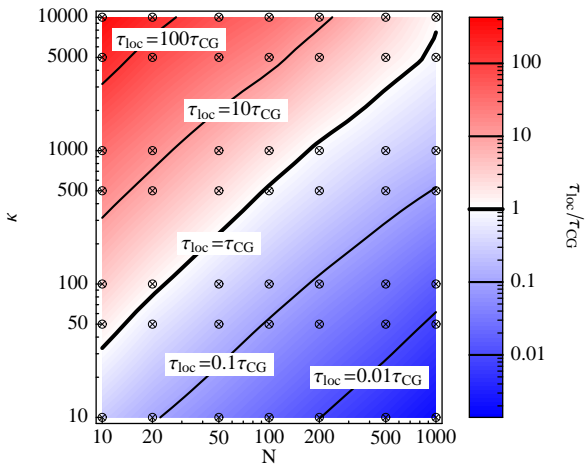}
\end{figure}

In Figure~\ref{fig:benchmark} we plot the ratios of the autocorrelation times 
$\tau\left[\Omega\right]$ as obtained with
local heatbath moves and with the block conjugate gradient version of 
our algorithm, as a function of changing condition number and system size.

\section{\label{sec:comp-global}Comparison with global heatbath}
It remains for us to discuss how our method fares in comparison with global heatbath.
The latter requires that matrix $\tens{\bf A}$ be decomposable in the 
form $\tens{\bf A}=\tens{\bf M}^T\tens{\bf M}$. This is the case in many
fields\cite{kraj-parr05prb}, but in principle if it were necessary to 
decompose $\tens{\bf A}$ this would add extra cost. 
Here we make our comparison assuming that $\tens{\bf M}$ is already
available. In such a case, the two algorithms are on paper equally efficient
in producing statistically independent samples. 
The global heatbath might offer some numerical advantages when
the spectrum of $\tens{\bf M}$ is highly degenerate, 
since the number of CG iterations needed to solve
the $\tens{\bf M}x={\bf R}$ linear system is $p<N$, as discussed earlier. 
Whenever a good preconditioner for the linear system is available, other 
inversion algorithms such as the stabilized bi-conjugate gradient\cite{vors92jssc}
or the generalized conjugate residual may allow to solve the linear system
with a sufficient accuracy more efficiently than using CG. In this
paper we make the comparison with conjugate gradient because of the 
close analogy with our scheme and because our method is aimed at problems
where ill-conditioning cannot be otherwise relieved.

\begin{figure}
\caption{\label{fig:omegatails}Autocorrelation function for the 
observable (\ref{eq:omega}) for an action of the form (\ref{eq:mat-phonon}),
with size $N=100$ and condition number $\kappa=5\times 10^3$.
Line {\bf A} corresponds to local heatbath, line {\bf B}
to the ``hybrid'' versions of our CG algorithm,
with a pool of two vectors with random restarts, while curve {\bf C} is obtained 
including the tricks described in section~\ref{sec:comp-global} 
with $m=5$}
\includegraphics[width=0.9\columnwidth]{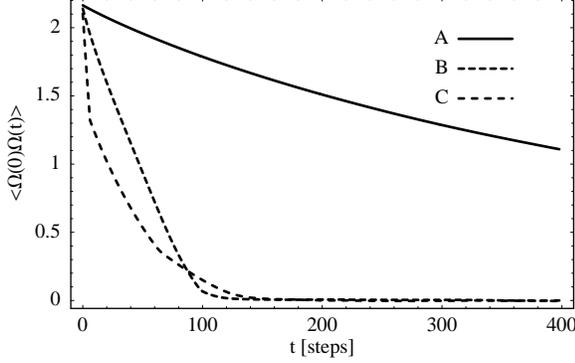}
\end{figure}

In this respect, our method displays significant advantages.
Firstly, it is more stable, because every move preserves the probability 
distribution, and the conjugate gradient procedure (which is known to be 
quite delicate in problems with large condition number) is only used
to generate search directions.
Instabilities in the procedure, which would cause incorrect sampling in the global
heatbath, affect only the efficiency, and not the accuracy.
Moreover, dividing the $N$ steps of an iterative inversion process into 
separate heatbath moves greatly improves the flexibility of the sampling scheme.
To give some examples, if one needs to perform an average on a slowly 
varying $\tens{\bf A}$, it is possible to perform only a partial sweep with 
fixed action, then continue with the new $\tens{\bf A}$, assuming that eigenmodes
will change slowly. It is also straightforward to tailor the choice of
directions in order to optimize the convergence speed for the observable
or interest. Adler's overrelaxation\cite{adle81prd} can be included
naturally, and can help in further optimizing the autocorrelation time.
 As an example of possible fine-tunings, 
let us recall the observable $\Omega$ introduced 
in the previous section (equation (\ref{eq:omega})). This observable depends 
strongly on the softer eigenvector of $\tens{\bf{A}}$. We have then modified 
our algorithm in the following way: we perform block conjugate gradient 
sweeps, with random resets, and we monitor the curvature along the direction being
thermalized, ${\bf h}\tens{\bf A}{\bf h}/{\bf h}\cdot{\bf h}$.
We save the direction of minimum curvature encountered along the sweep, 
${\bf h}_{min}$; during the following sweep, every $m$ 
moves along the CG directions, one move is performed along ${\bf h}_{min}$.
As is evident from Figure~\ref{fig:omegatails}, this trick considerably reduces the
autocorrelation time for $\Omega$. Even smarter combinations of moves can be 
devised, and the one we suggest is just an example of how the additional flexibility
gained through subdividing the inversion process in $N$ exact sampling
moves can be exploited.
In Table~\ref{tab:refvalues} we report some numerical extimates of the error in the
evaluation or $\Omega$, which can serve as a reference to compare our method
to other approaches.

\begin{table}
\caption{\label{tab:refvalues}
Percentual errors in the evaluation of $\Omega=\left<{\bf x^2}\right>$  (equation (\ref{eq:omega})),
extimated using a blocking analysis,  
for different sampling methods. {\bf A} corresponds to local heatbath, {\bf B} 
corresponds to ``hybrid'' versions of our CG algorithm,
with a pool of two vectors with random restarts, while curve {\bf C} is obtained 
including the tricks described in section~\ref{sec:comp-global} 
with $m=50$.
Different tests are performed with varying matrix size $N$, number of sampling steps $T$ 
and condition number $\kappa$. 
Due to the large autocorrelation time, the values of the error for local heatbath 
with $N=100$ and $T=10^6$ could not be extimated as reliably as in the other cases,
and are only indicative.}
\begin{ruledtabular}
\begin{tabular}[c]{c c c c c c}
  $N$  &   $\kappa$    &   $T$  & {\bf A} & {\bf B} & {\bf C} 	\\ \hline
$10^3$ &$5\times10^4$ & $10^6$ & 4.0  	 &	1.5	 &  1.4		\\ 
$10^3$ &$5\times10^4$ & $10^7$ & 1.3	 &	0.51	 &	0.45		\\ 
$10^3$ &$5\times10^3$ & $10^6$ & 0.78		 &	0.85   &  0.85	   	\\ 
$10^3$ &$5\times10^3$ & $10^7$ & 0.24	 &	0.28	 &	0.28  	\\ 
$100 $ &$5\times10^4$ & $10^6$ & $\sim$11 	 &	1.2	 &  1.1 		\\ 
$100 $ &$5\times10^4$ & $10^7$ & 4.9 	 &	0.44 &	0.34		\\ 
$100 $ &$5\times10^3$ & $10^6$ & $\sim$3   &	0.88  &  0.82 	   	\\ 
$100 $ &$5\times10^3$ & $10^7$ & 1.1	 &	0.30 &	0.25     	\\ 
\end{tabular}
\end{ruledtabular}
\end{table}

\section{\label{sec:conclusions}Conclusions}
We have presented an algorithm for performing collective modes heatbath along 
conjugate directions for a quadratic action, which allows  the components of
 the sampling vector along all modes to be decorrelated in $N$ steps, 
with a linear decay to zero.
 This method is  more computationally demanding than local updates, 
but  becomes competitive for ill-conditioned actions, when one needs to 
compute observables which depend on modes with low eigenvalues, or when the 
spectrum of the action matrix has only a few high eigenvalue modes which would 
slow down Cartesian moves.
 In fact, this method has an efficiency comparable with that of direct 
inversion of the matrix, but presents various advantages, such as 
improved stability, as the numerical issues connected with
conjugate gradient method do not affect the accuracy of the sampling, 
and the possibility of exploiting some additional
flexibility to improve the sampling on a case-by-case basis. Lastly, global
heatbath requires the knowledge of the square root of the action $\tens{\bf A}$,
so our scheme should be considered whenever the square root is difficult to 
compute or its use is inefficient with respect to the original action.

The geometrical simplicity of this approach, with its close analogy 
with minimization methods, also suggests that it might be extended 
to the sampling of anharmonic systems.

\appendix
\section{\label{sec:stationary}}
We report here a simple demonstration of the fact that heatbath moves
along a generic direction ${\bf d}$ leave an equilibrium probability
distribution unchanged. 
We will use the fact that if ${\bf R}$, ${\bf R}'$ and ${\bf R}''$ are vectors 
distributed as Gaussians with zero mean and standard deviation one, then
$\tens{\bf B}{\bf R}+\tens{\bf C}{\bf R}'$ is distributed as $\tens{\bf D}{\bf R}''$ 
where $\tens{\bf D}^T\tens{\bf D}=\tens{\bf B}^T\tens{\bf B}+\tens{\bf C}^T\tens{\bf C}$.
Since ${\bf x}$ is drawn from the equilibrium distribution, 
i.e. ${\bf x}=\tens{\bf M}^{-1}{\bf R}$,
we can cast Eq. (\ref{eq:hb-tau}) and (\ref{eq:step}) into the form
\begin{align}
x_j'=\sum_m P_{jm} R_m + \sum_m Q_{jm} R_m' \nonumber\\ 
P_{jm}=\left(\tens{\bf M}^{-1}\right)_{jm}-d_j \sum_k M_{km}d_k \quad\quad
Q_{jm}=d_j \delta_{m0}\nonumber
\end{align}
where we have put ${\bf b}=0$ into Eq. (\ref{eq:hb-tau}) and normalized
the direction so that ${\bf d}\tens{\bf A}{\bf d}=1$ in order to simplify the notation.
We can then compute
\[
\sum_m P_{jm}P_{lm}=\tens{\bf A}^{-1}_{jl}-d_j d_l \quad\quad
\sum_m Q_{jm}Q_{lm}=d_j d_l
\]
so that $\tens{\bf P}^T\tens{\bf P}+\tens{\bf Q}^T\tens{\bf Q}=
\left(\tens{\bf M}^{-1}\right)^T\tens{\bf M}^{-1}$, i.e. also ${\bf x'}$ may be  
written as $\tens{\bf M}^{-1}{\bf R}$, and is therefore correctly distributed.

\section{\label{sec:rnd-asymptotic}}
We shall here discuss briefly the derivation of the asymptotic form of equation 
(\ref{eq:random-slope}) when the size $N$ of the action matrix tends to infinity.
The quantity to be computed is 
\[
Q_i=\left<\frac{R_i^2}{\sum_k\eigena_k R_k^2}\right>\propto
\int\mathrm{d}{\bf x} \frac{x_i^2}{\sum_k x_k^2 \eigena_k} 
\exp\left[-\frac{1}{2}\sum_k x_k^2\right]
\]
The integral can be transformed as follows:
\begin{align*}
Q_i\propto\int_0^{\infty}\mathrm{d}t\int\mathrm{d}{\bf x} x_i^2
\exp\left[-\frac{1}{2}\sum_k \left(1+\eigena_k t\right)x_k^2\right]=\\
=\int_0^{\infty}\mathrm{d}t \frac{1}{\eigena_i t+1}\prod_k \frac{1}{\sqrt{\eigena_k t +1}},
\end{align*}
and the resulting expression, including the correct normalization, is
\begin{equation}
Q_i=\frac{1}{2}\int_0^{\infty}\mathrm{d}t \frac{1}{\eigena_i t+1} f\left(t\right),
\qquad f\left(t\right)=\prod_k \frac{1}{\sqrt{\eigena_k t +1}} \label{eq:asy-qi}
\end{equation}
Let us focus on $F=\int_0^{\infty}f\left(t\right){\rm d}t$, since all the $Q_i$ 
can be computed as 
$Q_i=\eigena_i \frac{\partial F}{\partial \eigena_i}+\frac{1}{2}F$. 
We perform the change 
of variables $Nt\rightarrow t$, so that 
\[
\int_0^{\infty}f\left(t\right){\rm d}t=
\frac{1}{N}\int_0^{\infty}\tilde{f}\left(t\right){\rm d}t,\qquad
\tilde{f}\left(t\right)=\prod_k \frac{1}{\sqrt{\frac{\eigena_k}{N} t +1}}.
\]
Under the physically reasonable assumption that $\Tr \tens{\bf A} = 
\mathcal{O}\left(N\right)$, and that the maximum eigenvalue 
does not scale with the system size, we can use $1/N$ as a small parameter.
Expanding $\log\tilde{f}$ one finds
\begin{eqnarray*}
\log\tilde{f}\left(t\right)=\sum_k \log\left(1+\frac{\eigena_k}{N}t\right)=\\
\sum_{n=1} \frac{t^n}{n+1} \sum_k \left[\frac{\eigena_k}{N}\right]^n
=\sum_k\frac{\eigena_k}{N}\frac{t}{2} + \sum_{n=1} t^{n+1} \mathcal{O}\left(\frac{1}{N^n}\right).
\end{eqnarray*}
All but the leading term become negligible for $N\rightarrow \infty$.
This suggests separating out from $\tilde{f}\left(t\right)$ the term order zero
in $1/N$, and writing for $F$ the expression 
\begin{eqnarray}
\frac{1}{N}\int_0^{\infty}\exp\left(-\frac{t}{2}\frac{\Tr \tens{\bf A}}{N}\right)\times
\nonumber\\
\times\left[1+\frac{1}{4}\sum_k\left(\frac{\eigena_k}{N}\right)^2 t^2 + 
\mathcal{O}\left(\frac{1}{N^2}\right)t^3+\ldots
\right]
{\rm d} t
\end{eqnarray}
which leads to the asymptotic result $F=\frac{2}{Tr\tens{\bf A}}+\mathcal{O}\left(N^{-2}\right)$.
Correspondingly, dropping the higher order terms in $1/N$, we have
$Q_i=\frac{1}{Tr\tens{\bf A}}+\mathcal{O}\left(N^{-2}\right)$, which is the
desired  result.

\section{\label{sec:cd-formal}}
We obtain here the autocorrelation function for the components along the 
eigenmodes of the action matrix $\tens{\bf A}$, when performing heatbath
sweeps along a set of conjugate directions
 $\left\{{\bf h}^{(m)}\right\}_{m=0\ldots N-1}$. In this
section, the indices of the directions are defined modulo $N$, i.e.
${\bf h}^{(j+N)}={\bf h}^{(j)}$.
In this case, one can write Eq. (\ref{eq:auto-induction}) as 
\begin{align}
\left<x_i\left(0\right) x_i\left(t+1\right)\right>=
\left<x_i\left(0\right) x_i\left(t\right)\right>-\nonumber\\
\frac{1}{N}\sum_m\sum_k\left[\left<x_i\left(0\right) x_k\left(t\right)\right>
\sqrt{\frac{\eigena_k}{\eigena_i}}\Delta_{ki}\left({\bf h}^{(m)}\right)\right].
\label{eq:cg-induction}
\end{align}
Explicit calculations for small values of $t$ suggest for $t<N$ the ansatz 
\begin{equation}
\left<x_i\left(0\right) x_i\left(t\right)\right>=
\left<x_i\left(0\right)^2\right>
\left[1-\frac{t}{N}\right].\label{eq:cg-ansatz}
\end{equation}
Since the first term in Eq. (\ref{eq:cg-induction}) does not contain the new
direction, we can substitute the ansatz without concern. On the other hand, 
the second term contains reference to ${\bf h}^{(m)}$, so that the average 
that led to (\ref{eq:cg-ansatz}) cannot be performed separately, and one 
should rather write:
\begin{align}
\frac{1}{N}\sum_m\sum_k\left[
\sum_{k'}\left<x_i\left(0\right) x_{k'}\left(t-1\right)\right>\right.
\nonumber\\
\left.\left(\delta_{k'k}-\sqrt{\frac{\eigena_{k'}}{\eigena_k}}
\Delta_{k'k}\left({\bf h}^{(m-1)}\right)\right)
\sqrt{\frac{\eigena_k}{\eigena_i}}
\Delta_{ki}\left({\bf h}^{(m)}\right)\right].
\end{align}
which is split into
\begin{align}
\frac{1}{N}\sum_m\sum_k\left[\left<x_i\left(0\right) x_k\left(t-1\right)\right>
\sqrt{\frac{\eigena_k}{\eigena_i}}\Delta_{ki}\left({\bf h}^{(m)}\right)\right],
\label{eq:cg-one}\\
\frac{1}{N}\sum_{mkk'}\left[\left<x_i\left(0\right) x_{k'}\left(t-1\right)\right>
\sqrt{\frac{\eigena_{k'}}{\eigena_i}}
\Delta_{k'k}\left({\bf h}^{(m-1)}\right)
\Delta_{ki}\left({\bf h}^{(m)}\right)\right]\label{eq:cg-two}
\end{align}
The term (\ref{eq:cg-two}) goes to zero, since
\[
\sum_k\sum_m \Delta_{ik}\left({\bf h}^{(m-n)}\right)
\Delta_{kj}\left({\bf h}^{(m)}\right)=\delta_{n,pN}\delta_{ij}
\]
while (\ref{eq:cg-one}) can be expanded again, giving rise to the $t-2$ analogue and
to a term containing $\Delta_{k'k}\left({\bf h}^{(m-2)}\right)\Delta_{kj}
\left({\bf h}^{(m)}\right)$. One iterates this process recursively until it reaches 
$\left<x_i\left(0\right)^2\right>$, thus contributing another $-1/N$ to the 
autocorrelation function.
Things are different for $t\ge N$, since terms involving products of
the slopes for the same direction will enter the procedure at a certain point
in the iteration.
Because of these terms, for $t\ge N$ autocorrelation functions will be 
identically zero.

\section*{Acknowledgments}
It is a pleasure to acknowledge useful discussion with Fulvio Ricci and 
Nazario Tantalo, whose suggestions have helped improving the manuscript.

\end{document}